\documentclass[aps,prl,twocolumn,superscriptaddress]{revtex4-1}
\usepackage{bm}
\usepackage{graphicx}
\usepackage{color}
\usepackage{braket}
\usepackage{amsmath,amssymb,amsfonts,amsthm,mathtools}
\usepackage{enumerate}
\usepackage{enumitem}
\usepackage[colorlinks=true,linkcolor=blue,anchorcolor=red,citecolor=blue,urlcolor=blue]{hyperref}
\usepackage{diagbox}
\usepackage{titlesec}
\usepackage{tikz-cd}
\usepackage{float}
\usepackage{multirow}
\usepackage[title]{appendix}
\usepackage{pdfpages}
\usepackage{changes}
\usepackage{stackengine}
\usepackage{makecell}

\makeatletter
\AtBeginDocument{\let\LS@rot\@undefined}
\makeatother

\makeatletter 
\renewcommand\NAT@biblabelnum[1]{#1.} 
\makeatother

\setcitestyle{super}

\begin{document}

\title{Chern Insulators on a Twisted Klein Bottle}

\author{Rong Xiao}
\affiliation{Department of Physics and HK Institute of Quantum Science \& Technology, The University of Hong Kong, Pokfulam Road, Hong Kong, China}

\author{Y. X. Zhao}
\email[]{yuxinphy@hku.hk}
\affiliation{Department of Physics and HK Institute of Quantum Science \& Technology, The University of Hong Kong, Pokfulam Road, Hong Kong, China}

\begin{abstract}
Recently, momentum-space nonsymmorphic symmetries have attracted considerable attention. However, their realizations typically rely on fine-tuning hopping phases to satisfy the required projective symmetry algebras. Here, we show that such symmetry algebras arise naturally on bipartite lattices through the interplay of crystalline and sublattice symmetries. Unlike previously studied cases, these symmetry operators can anticommute with the Hamiltonian and are therefore referred to as momentum-space nonsymmorphic chiral symmetries. Although the free actions of these symmetries reduce the Brillouin torus to more elementary manifolds, the resulting manifolds are twisted in the Atiyah--Segal formalism. In particular, chiral glide reflection (screw rotation) gives rise to a twisted Klein bottle (dicosm) in two (three) dimensions, and this twisting dramatically alters the topological classification. Unlike the untwisted Klein bottle, whose nonorientability forces the Chern number to vanish, the twisted Klein bottle can host nonzero Chern numbers, corresponding to even Chern numbers on the torus. At open boundaries, the corresponding topological invariant manifests itself through chiral edge states exhibiting a glide-reflection structure in energy--momentum space. Our results establish momentum-space nonsymmorphic chiral symmetry as a new organizing principle for discovering and engineering topological phases beyond conventional crystalline classifications.
\end{abstract}

\maketitle

\noindent \textbf{INTRODUCTION}\\

\begin{figure*}[tb]
	\centering
	\includegraphics[width=2.0\columnwidth]{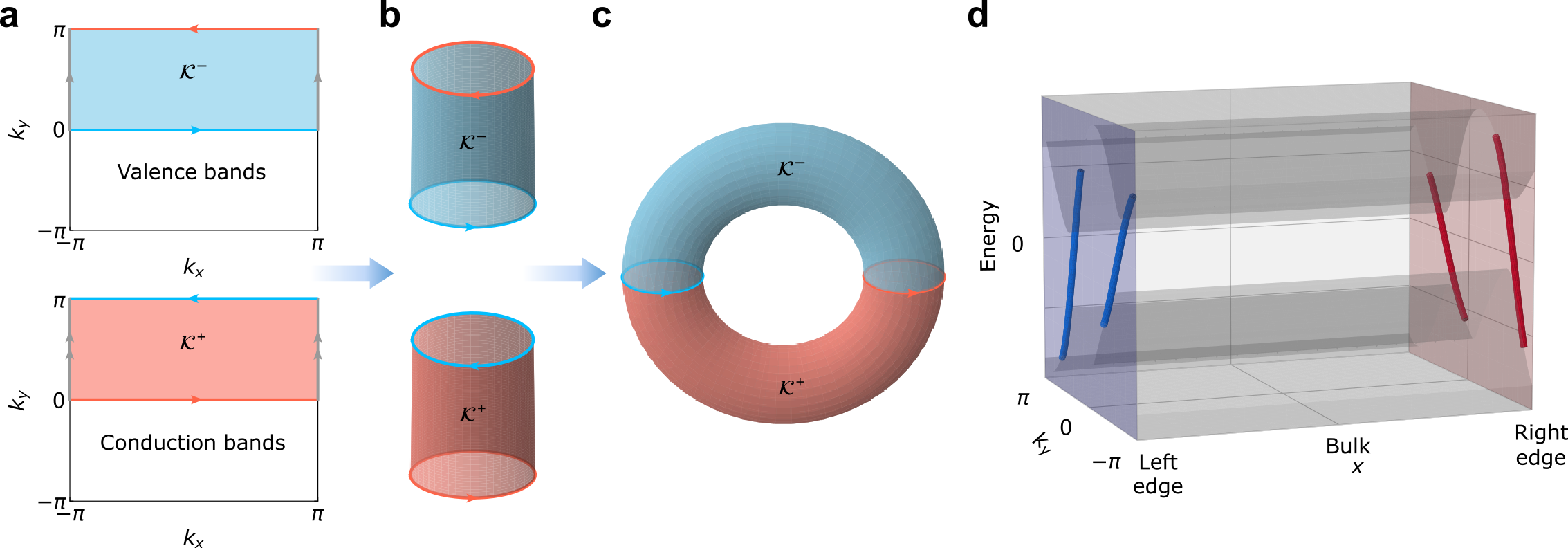}
	\caption{\textbf{Momentum space representation of the symmetry and twisted Brillouin
			Klein bottle.}
		\textbf{a} The valence and conduction Brillouin zones and the corresponding fundamental domains $\mathcal{K}^{\pm}$. \textbf{b} Cut the valence and conduction Klein bottles into cylinders and edge gluing rules specified by the twisting. \textbf{c} The torus resulting from gluing edges with the same color along the indicated directions. \textbf{d} Schematic illustration of edge spectra of the system in a slab geometry with $C=2$. States on the left/right boundary are marked in blue/red color. }
	\label{fig:topology}
\end{figure*}

Extending the symmetry classes of condensed-matter systems is important within the framework of symmetry-protected topological phases, as demonstrated by the tenfold classification of topological insulators and superconductors~\cite{Altland_1997,Schnyder_2008,Kitaev_2009,Ryu_2010,Chiu_2016}. When spatial symmetries are included, nonsymmorphic crystal symmetries, such as glide reflections and screw rotations~\cite{Bradley_2009}, can lead to intrinsic band crossings and thereby protect exotic topological phases~\cite{Michel_1999,Parameswaran_2013,Yang_2014,Young_2015,Watanabe_2015,Fang_2015,Shiozaki_2015,Shiozaki_2016,Zhao_2016,Wang_2016,Chang_2017}, in contrast to symmorphic symmetries. Recently, the concept of nonsymmorphic symmetry has been generalized to momentum space~\cite{Chen_2022,Zhang_2023,Zhang_2025} and realized in various artificial crystals~\cite{Zhu_2024,Vaidya_2025,Pu_2023,Fonseca_2024,Hu_2024,Lai_2024,Tao_2024,Liu2024,Hu_2025,Qiu_2025,Shen_2025,Li_2026} and condensed-matter systems~\cite{Xiao_2024Spin,Cualuguaru_2025,Yoon_2026}. Stemming from projective representations of crystal symmetries, momentum-space nonsymmorphic symmetries can, through their free actions, reduce the Brillouin torus to a Klein bottle in two dimensions~\cite{Chen_2022} and to any nontrivial platycosm in three dimensions~\cite{Zhang_2025}. Topological phases protected by momentum-space nonsymmorphic symmetries have been theoretically classified~\cite{Chen_2022,Zhang_2025} and experimentally explored in various artificial crystals~\cite{Pu_2023,Fonseca_2024,Hu_2024,Lai_2024,Tao_2024,Liu2024,Hu_2025,Qiu_2025,Shen_2025,Li_2026}.

However, constructing the relevant projective representations usually requires fine-tuning the phases of hopping amplitudes~\cite{Chen_2023}, which limits their realization in more general contexts. In this work, we present a universal mechanism that utilizes the recently revealed spatial nature of sublattice symmetry $\Gamma$ on bipartite lattices~\cite{Xiao_2024}, i.e., whether a crystal symmetry preserves or reverses the bipartition of the lattice. The composite operator $g\Gamma$, which combines a spatial symmetry $g$ with the sublattice symmetry $\Gamma$, can naturally satisfy the projective symmetry algebra required for momentum-space nonsymmorphic symmetries. This purely symmetry-based approach depends only on the lattice geometry and therefore fundamentally enhances the realizability of momentum-space nonsymmorphic symmetries.

Crucially, because the momentum-space nonsymmorphic operator inherits the anticommuting nature of the sublattice symmetry, it anticommutes with the Hamiltonian. This defines a new type of symmetry, termed \textit{momentum-space nonsymmorphic chiral symmetry}.

Analogous to previously studied momentum-space symmetries, momentum-space nonsymmorphic chiral symmetries can, through their free actions, reduce the Brillouin torus to an orbit manifold $\mathcal{M}$, such as a Klein bottle for a glide reflection. However, chiral symmetries exchange the valence and conduction bands, leading to a twisting $\tau\in H^1(\mathcal{M}, \mathbb{Z}_2)$ on $\mathcal{M}$ in the Atiyah--Segal formalism~\cite{Atiyah_2004}. The topological classification on the \textit{$\tau$-twisted $\mathcal{M}$}, denoted by $\mathcal{M}_\tau$, is given by the twisted K-group $K(\mathcal{M}_\tau)$, which can be completely different from the topological classification $K(\mathcal{M})$ on the untwisted $\mathcal{M}$.

In this work, we construct a chiral glide-reflection symmetry and a chiral twofold screw-rotation symmetry in momentum space. In particular, whereas the Chern number is always trivial on the untwisted Klein bottle obtained from a glide reflection and in standard class AIII, the twisted Klein bottle can host arbitrary integer Chern numbers, and therefore the Chern number over the entire Brillouin torus is always an even integer.  In the unit case, the two chiral modes on one edge are mapped to the two on the other edge by an energy--momentum glide reflection on a slab geometry. For the twofold screw rotation in $3$D, the topological classification over the twisted platycosm is characterized by two Chern numbers defined on the two embedded twisted sub-Klein bottles.

We construct concrete lattice models to demonstrate the unconventional topological features arising from the twisting associated with momentum-space nonsymmorphic chiral symmetries.

\bigskip
\noindent \textbf{RESULTS}\\
\textbf{Momentum-space nonsymmorphic chiral symmetries}.
There are two types of nonsymmorphic symmetries: glide reflections and screw rotations~\cite{Bradley_2009}. We begin by considering momentum-space glide reflection symmetry. As shown in Ref.~\cite{Chen_2022}, this momentum-space symmetry arises from the real-space projective symmetry algebra $\{M_x,T_y\}=0$, or equivalently, $M_xT_yM_x^{-1}=-T_y$, where $M_x$ is the mirror reflection that inverts the $x$ coordinate and $T_y$ is the primitive lattice translation by $b$ along the $y$ direction. In momentum space, $T_y$ is represented by $e^{i k_y b}$, and therefore the minus sign on the right-hand side manifests itself as a half-reciprocal-lattice translation ${G}_y/2$.

Previously, the projective algebra was realized by appropriately modulating gauge fluxes on the lattice, i.e., by fine-tuning the hopping phases. Here, however, we consider the sublattice operator $\Gamma$ of a bipartite lattice. The operator $\Gamma$ is a gauge transformation that assigns opposite signs to the two sublattices. On a bipartite lattice, a spatial transformation $g$ either preserves or reverses the bipartition and, correspondingly, commutes or anticommutes with $\Gamma$. Thus, when $T_y$ reverses the bipartition, $\Gamma$ translates $k_y$ by $G_y/2$ in momentum space~\cite{Xiao_2024}. To construct a momentum-space glide reflection, we need only preserve the combined symmetry $M_x\Gamma$, while the two individual symmetries may be broken. In practice, we first construct a Hamiltonian $\mathcal{H}_0$ that preserves both $M_x$ and $\Gamma$ and then add a perturbation $V$ that breaks both symmetries while preserving their combination, leading to a perturbed Hamiltonian $\mathcal{H}=\mathcal{H}_0+V$.

\begin{figure}[tb]
	\centering
	\includegraphics[width=\columnwidth]{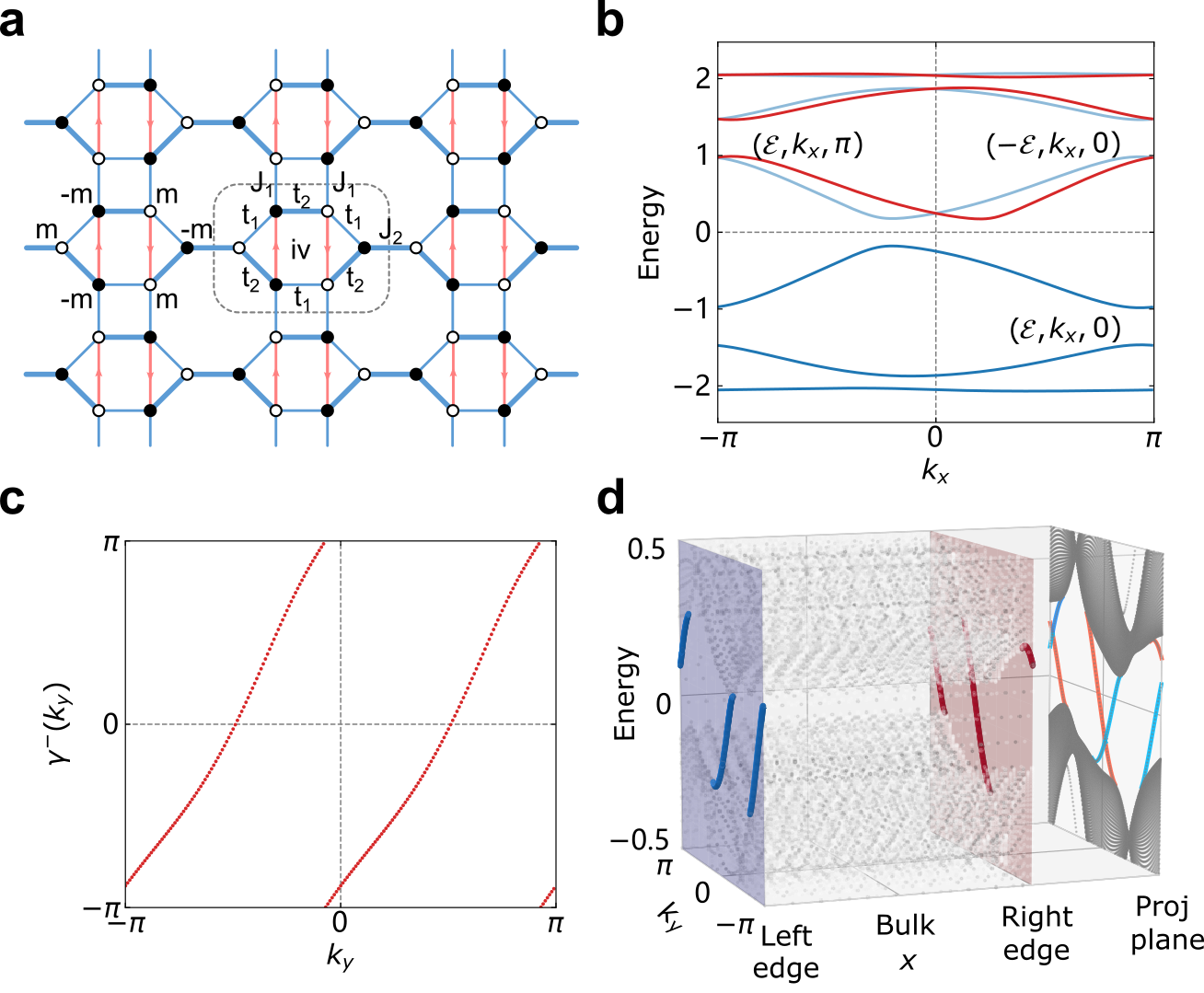}
	\caption{
		\textbf{Even Chern number and unconventional chiral edge states.} 
		\textbf{a} Schematic of the tight-binding model on the two-dimensional SHO lattice. The SHO lattice bonds are shown in blue, the imaginary hopping perturbations are shown in red, and \(\pm m\) denote the on-site potentials.
		\textbf{b} Cross sections of the band structure at \(k_y=0\) and \(k_y=\pi\), illustrating the screw-rotation symmetry in energy--momentum space.
		\textbf{c} Berry-phase flow \(\gamma^-(k_y)\).
		\textbf{d} Energy spectrum of the model in a slab geometry. The chiral states localized at the two boundaries are shown in red and blue, respectively. 
	}
	\label{fig:sho-model}
\end{figure}

Although the resulting symmetry $M_x\Gamma$ acts as a glide reflection in momentum space,
\begin{equation}
	M_x\Gamma: (k_x,k_y) \mapsto (-k_x,k_y+G_y/2),
\end{equation}
it anticommutes with the Hamiltonian,
\begin{equation} \label{eq:symmetry-algebra}
	U_{M_x\Gamma} \mathcal{H}(k_x,k_y) U_{M_x\Gamma}^{\dagger}=-\mathcal{H}(-k_x,k_y+G_y/2),
\end{equation}
where $U_{M_x\Gamma}$ is a unitary operator. Thus, this symmetry should be called a momentum-space glide-reflection chiral symmetry and is fundamentally different from the previously introduced momentum-space glide reflection symmetry. Its action can also be viewed in energy--momentum space as
\begin{equation}
	M_x\Gamma: (\mathcal{E},k_x,k_y)\mapsto (-\mathcal{E},-k_x,k_y+G_y/2).
\end{equation}
For each eigenstate $|u(\bm{k})\rangle$ with energy $\mathcal{E}(\bm{k})$, Eq.~(\ref{eq:symmetry-algebra}) requires the transformed state $U_{M_x\Gamma}|u(\bm{k})\rangle$ to satisfy
$\mathcal{H}(-k_x,k_y+G_y/2) U_{M_x\Gamma}|u(\bm{k})\rangle=-\mathcal{E}(\bm{k})U_{M_x\Gamma}|u(\bm{k})\rangle$.
Thus, the band structure features a screw rotation symmetry in the $(\mathcal{E},\bm{k})$ space [see an example in Fig.~\ref{fig:sho-model}b].


We now turn to screw rotations. Since our construction relies on interpreting the sublattice operator as a $\mathbb{Z}_2$ gauge transformation on a bipartite lattice, this mechanism can only produce a factor system valued in $\{\pm 1\}$ and, consequently, only half-reciprocal-lattice translations. Therefore, the only momentum-space screw rotation that can arise from this mechanism is a twofold screw rotation.

The simplest real-space group for realizing this symmetry is $P2$, generated by three primitive translations $T_{x,y,z}$ and a twofold rotation $R$ about the $z$ axis. If $\{\Gamma, T_z\}=0$ and $[\Gamma, T_{x,y}]=0$, then $(R\Gamma) T_z (R\Gamma)^{-1}=-T_z$ and $(R\Gamma) T_{x,y} (R\Gamma)^{-1}=T_{x,y}^{-1}$. Thus, $R\Gamma$ acts on momentum space precisely as a twofold screw rotation
\begin{equation}
	R\Gamma:(k_x,k_y,k_z)\mapsto (-k_x,-k_y,k_z+G_z/2).
\end{equation}
Similarly, for the $R\Gamma$-preserving Hamiltonian, the symmetry constraint is
\begin{equation}
	U_{R\Gamma} \mathcal{H}(k_x,k_y,k_z) U_{R\Gamma}^{\dagger}=-\mathcal{H}(-k_x,-k_y,k_z+G_z/2),
\end{equation}
where $U_{R\Gamma}$ is a unitary operator. Consequently, $R\Gamma$ acts on the energy--momentum space as
\begin{equation}
	R\Gamma: (\mathcal{E},k_x,k_y,k_z)\mapsto (-\mathcal{E},-k_x,-k_y,k_z+G_z/2).
\end{equation}
Hence, $R\Gamma$ inverts the energy $\mathcal{E}$ and the momenta $k_x$ and $k_y$ while simultaneously translating $k_z$ by $G_z/2$.

\smallskip

\noindent\textbf{Twisting altered topological classification}.
As in the previous cases, the free actions of nonsymmorphic symmetries reduce the Brillouin torus to an orbit manifold $\mathcal{M}$, which is a Klein bottle for the glide reflection and the platycosm known as the dicosm for the twofold screw rotation. The group action dictates how the boundary components of the fundamental domain are identified to form the reduced manifold. The essential difference, however, is that momentum-space nonsymmorphic chiral symmetries exchange the valence and conduction bands under this identification. Whether the valence and conduction bands are exchanged or preserved can be encoded by an element $\tau$ of $H^1(\mathcal{M},\mathbb{Z}_2)$, called a twisting of the action~\cite{Freed_2013}.

To see the geometric meaning of $\mathcal{K}_\tau$, we may introduce Brillouin zones for both the valence and conduction bands [see Fig.~\ref{fig:topology}a] and represent the Klein bottle as a fundamental domain on each of them. Each Klein bottle results from gluing the two edges of a cylinder along opposite directions [see Fig.~\ref{fig:topology}b]. Then, the twisting $\tau$ can be regarded as a new gluing rule for the two cylinders, i.e., the lower (upper) edge of $C_{-}$ ($C_{+}$) should be glued with the upper edge of $C_{+}$ ($C_{-}$) along the direction indicated in Fig.~\ref{fig:topology}b. The resulting manifold is a torus connecting $C_{-}$ and $C_{+}$ [see Fig.~\ref{fig:topology}c]. Mathematically, the resulting torus is a $\mathbb{Z}_2$ bundle over $\mathcal{K}$ determined by $\tau$.

The twisted free action of momentum-space nonsymmorphic symmetries profoundly alters the topological classification. To determine this classification, we compute the twisted K-group $K(\mathcal{M}_\tau)$ using the Atiyah--Hirzebruch spectral sequence~\cite{Atiyah_1961}. For the Klein bottle $\mathcal{K}$, it is useful to formulate the twisting $\tau\in H^1(\mathcal{K},\mathbb{Z}_2)$ algebraically in order to describe the second page of the spectral sequence. Because $\mathcal{K}=\mathbb{R}^2/pg$ under the natural action of the wallpaper group $pg$, we have $H^1(\mathcal{K},\mathbb{Z}_2)\cong H^1(pg,\mathbb{Z}_2)$. A twisting can therefore be interpreted as a homomorphism from $pg$ to $\mathbb{Z}_2$, which in the present case is given by $\tau(T_x)=0$ and $\tau(G_x)=1$. Here, $pg$ is generated by the lattice translation $T_x$ along the $x$ direction and the glide reflection $G_x$ that inverts the $x$ coordinate. Accordingly, $\tau$ specifies an action of $pg$ on $\mathbb{Z}$. The second page can thus be written as $E^{p,q}_2=H^p(\mathcal{K},\mathbb{Z}_\tau)\cong H^p(pg,\mathbb{Z}_\tau)$ for even $q$, and $E^{p,q}_2=0$ for odd $q$. Since the only nontrivial cohomology groups are $H^1(pg,\mathbb{Z}_\tau)\cong\mathbb{Z}\oplus \mathbb{Z}_2$ and $H^2(pg,\mathbb{Z}_\tau)\cong\mathbb{Z}$, the second page is already stable, and we conclude that
\begin{equation}\label{eq:twisted-K}
	K(\mathcal{K}_\tau)\cong\mathbb{Z}.
\end{equation}
The method of expressing the second page in terms of group cohomology also applies to the twofold screw rotation associated with the space group $P2_{1}$. This group is generated by the translations $T_x$ and $T_y$ and the screw rotation $S_z$. Only $S_z$ inverts $\mathbb{Z}$, thereby specifying the corresponding action of $P2_1$ on $\mathbb{Z}$. The nontrivial cohomology groups are $H^1(P2_1,\mathbb{Z}_\tau)\cong\mathbb{Z}^2\oplus \mathbb{Z}_2$, $H^2(P2_1,\mathbb{Z}_\tau)\cong \mathbb{Z}^2$, and $H^3(P2_1,\mathbb{Z}_\tau)\cong \mathbb{Z}_2$; therefore, the second page is also stable. Consequently, the classification is $K(\mathcal{D}i_\tau)\cong \mathbb{Z}\oplus\mathbb{Z}$, where $\mathcal{D}i\cong \mathbb{R}^3/P2_1$. The two integer components can be readily understood by observing that $\mathcal{D}i$ contains two embedded sub-Klein bottles, each of which contributes a $\mathbb{Z}$ according to $K(\mathcal{K}_\tau)\cong\mathbb{Z}$.

\smallskip
	
\noindent\textbf{Even Chern number}.
It is now clear that the central problem in characterizing the two topological classifications is to formulate a topological invariant on the twisted Klein bottle or, equivalently, on the Brillouin torus with the twisted glide-reflection symmetry. In the following, we show that, instead of forcing the Chern number to vanish as in the preceding cases of class AIII~\cite{Ryu_2010} and the untwisted Klein bottle~\cite{Chen_2022}, the twisted glide-reflection symmetry constrains the Chern number to be an even integer.

Let \(|u^{\pm}_n(\bm{k})\rangle\) be orthonormal bases for the conduction and valence bands, respectively, of a symmetry-preserving Hamiltonian \(\mathcal{H}(\bm{k})\) gapped at zero energy. The corresponding Abelian Berry connections are $a^{\pm}_{\mu}(\bm{k})=\sum_n\langle u^{\pm}_{n}(\bm{k})|i\partial_{k_\mu}|u^{\pm}_{n}(\bm{k})\rangle,$
and the Berry curvatures are
$f^{\pm}(\bm{k})=\partial_{k_x}a^\pm_y-\partial_{k_y}a^\pm_x$.
We define \(\Phi[a,b]\) as the flux of \(f^-(\bm{k})\) through the momentum-space region \([-\pi,\pi]\times[a,b]\), that is,
$\Phi[a,b]=\int_a^b dk_y\oint dk_x\,f^-(\bm{k})$.
We can always choose \(|u^{\pm}_{n}(\bm{k})\rangle\) to be periodic along the \(k_x\) direction~\cite{Thouless_1983}. Let \(h_{k_y}(k_x)=\mathcal{H}(\bm{k})\) denote the Hamiltonian of the \(k_x\) subsystem parametrized by \(k_y\). We then define the Berry phases of \(h_{k_y}(k_x)\) as
$\gamma^{\pm}(k_y)=\oint dk_x\,a^{\pm}_{x}(\bm{k})$.
It follows directly that
\[
\Phi[a,b]=\gamma^-(a)-\gamma^-(b).
\]
Hereafter, for simplicity, we assume that every reciprocal primitive lattice vector has length $2\pi$.
For an arbitrary fundamental domain \([-\pi,\pi]\times[k_y,k_y+\pi]\) of the glide reflection, we show that
\begin{equation}\label{eq:Chern_K}
	C_{\mathcal{K}}
	=
	\frac{\Phi[k_y,k_y+\pi]}{2\pi}
	\in\mathbb{Z}.
\end{equation}
Thus, \(C_{\mathcal{K}}\) is an integer-valued invariant of the twisted Klein bottle and characterizes the classification in Eq.~\eqref{eq:twisted-K}.

To establish Eq.~\eqref{eq:Chern_K}, we observe that \(U_{M_x\Gamma}|u_n^{-}(\bm{k})\rangle\) is a positive-energy eigenstate of \(\mathcal{H}(-k_x,k_y+\pi)\), as follows from Eq.~\eqref{eq:symmetry-algebra}. Therefore,
\begin{equation}
	U_{M_x\Gamma}|u_n^{-}(\bm{k})\rangle
	=\sum_m
	|u_m^{+}(-k_x,k_y+\pi)\rangle
	\mathcal{U}_{mn}(\bm{k}),
\end{equation}
where \(\mathcal{U}(\bm{k})\) is unitary and periodic in \(k_x\). The Berry connections \(a_x^\pm\) are consequently related by
\begin{equation}
	a_x^{-}(\bm{k})
	=-a_x^{+}(-k_x,k_y+\pi)
	+i\operatorname{Tr}
	\left[
	\mathcal{U}^\dagger(\bm{k})
	\partial_{k_x}\mathcal{U}(\bm{k})
	\right].
\end{equation}
This gives the relation
\begin{equation}
	\gamma^-(k_y)
	=-\gamma^+(k_y+\pi)
	\pmod{2\pi}.
\end{equation}
Note that
$\frac{i}{2\pi}
\int_{-\pi}^{\pi}dk_x\,
\operatorname{Tr}
\left[
\mathcal{U}^\dagger(\bm{k})
\partial_{k_x}\mathcal{U}(\bm{k})
\right]$
is the winding number of \(\mathcal{U}(k_x,k_y)\) as \(k_x\) traverses \([-\pi,\pi]\), and is therefore a constant independent of \(k_y\). For each \(k_y\), \(\gamma^+\) and \(\gamma^-\) are opposite modulo a large gauge transformation, so that
\begin{equation}
	\gamma^-(k_y+\pi)
	=-\gamma^+(k_y+\pi)
	\pmod{2\pi}.
\end{equation}
Equation~\eqref{eq:Chern_K} then follows from these two identities.

Because the flux through any fundamental domain is quantized in units of \(2\pi\), the flux through the entire Brillouin zone is an even integer multiple of \(2\pi\). Consequently, the Chern number \(C\) must be even in the presence of chiral glide-reflection symmetry in momentum space. The minimal nontrivial case is \(C=2\), for which each edge of a slab hosts two chiral modes. When the system is confined along the \(x\) direction, \(M_x\Gamma\) relates the band structures on the two edges. Under
\begin{equation}
	(\mathcal{E},k_y)
	\longmapsto
	(-\mathcal{E},k_y+\pi),
\end{equation}
the two chiral bands on one edge are mapped to those on the opposite edge [see Fig.~\ref{fig:topology}d]. In other words, the two edge spectra are related by a glide reflection in energy--momentum space.

\smallskip

\noindent\textbf{\(2\)D lattice model with chiral glide reflection}.
We now demonstrate momentum-space chiral glide-reflection symmetry using a tight-binding model defined on the two-dimensional square--hexagon--octagon (SHO) lattice, whose bonds are shown in blue in Fig.~\ref{fig:sho-model}a. Here, we only briefly present the main features of the model, and further details can be found in Methods.

On the SHO lattice, the unperturbed Hamiltonian that only contains nearest-neighbor hoppings is $\mathcal{H}_0$, whose explicit form is provided in Methods. The Hamiltonian \(\mathcal{H}_0(\bm{k})\) preserves the sublattice symmetry \(\Gamma=\tau_3\otimes 1_3\,\mathcal{L}_{\bm{G}_y/2}\) and the \(Pm\) symmetry group generated by the reflection \(M_x=\tau_1\otimes 1_3\,\hat{m}_x\). To explicitly break \(\Gamma\) and \(M_x\) while preserving their product \(M_x\Gamma\), we introduce the perturbation \(V\), whose explicit form is also given in Methods. Thus, the total Hamiltonian is given by \(\mathcal{H}(\bm{k})=\mathcal{H}_0(\bm{k})+V\), as illustrated in Fig.~\ref{fig:sho-model}a. Note that time-reversal symmetry must also be broken for the Chern number to be nonzero.

The bulk band structure of \(\mathcal{H}(\bm{k})\) exhibits the predicted screw-rotation symmetry in \((\mathcal{E},\bm{k})\) space. For example, as shown in Fig.~\ref{fig:sho-model}b, the energy-inverted cross section of the valence-band structure at \(k_y=0\) coincides with the mirror reflection of the conduction-band cross section at \(k_y=\pi\).

Figure~\ref{fig:sho-model}c shows the Berry phase \(\gamma^-(k_y)\) as a function of \(k_y\). The Berry-phase flow is \(\pi\)-periodic in \(k_y\) and completes two full windings over one \(2\pi\) period, yielding a Chern number \(C=2\). Figure~\ref{fig:sho-model}d shows the energy spectrum of the model in a slab geometry invariant under \(M_x\Gamma\). We observe two chiral bands on each edge traversing the bulk energy gap. Notably, because \(M_x\Gamma\) maps one edge to the other, the chiral-edge-state spectra on the two edges exhibit an unusual glide-reflection symmetry in \((\mathcal{E},k_y)\) space: the two edge spectra are exchanged by a $\pi$-translation along the \(k_y\) direction combined with energy inversion.

In addition, a three-dimensional lattice model with the chiral screw-rotation symmetry is presented in Supplementary Note 1.

\bigskip
	
\noindent\textbf{DISCUSSION}\\
In summary, we have established a symmetry-based framework for constructing momentum-space nonsymmorphic chiral symmetries on bipartite lattices. By combining sublattice symmetry with a crystalline reflection or twofold rotation, we obtain composite symmetries that anticommute with the Hamiltonian and act freely in momentum space as a glide reflection or screw rotation, without requiring fine-tuned hopping phases. These free actions reduce the Brillouin torus to a Klein bottle in two dimensions and a dicosm in three dimensions, while the associated exchange of valence and conduction bands introduces a nontrivial twisting that fundamentally alters the topological classifications.

The symmetry-based mechanism for realizing twisted projective symmetry algebras can be generalized to a broader context. In general, we can consider a space group $G$ with a $\mathbb{Z}_2$ grading $\alpha$, where $\alpha(g)=\pm 1$ specifies whether $g$ preserves or reverses the two sublattices. The total symmetry group $G\times \mathbb{Z}_2^\Gamma$, with $\mathbb{Z}_2^\Gamma=\{1,\Gamma\}$, is then reduced to a subgroup $\widetilde{G}_0$ specified by a subgroup $G_0$ of $G$ and a homomorphism $c$ from $G_0$ to $\mathbb{Z}_2^\Gamma$. Namely, $\widetilde{G}_0$ consists of $(g,c(g))$ for all $g\in G_0$. The grading $\alpha$ determines a factor system of $\widetilde{G}_0$, while $c$ specifies whether the symmetries in $\widetilde{G}_0$ commute or anticommute with the Hamiltonian.

\bigskip
\noindent \textbf{\large METHODS}\\

\noindent \textbf{Details of the 2D lattice model in Fig.~2}. 
We now present more details about the lattice model defined on the SHO lattice. 

The unperturbed Hamiltonian on the SHO lattice, containing only nearest-neighbor hoppings, takes the form
\begin{equation}
	\mathcal{H}_0(\bm{k})
	=
	\begin{bmatrix}
		p(k_y) & q(k_x)\\
		q^{\dagger}(k_x) & p(k_y)
	\end{bmatrix},
\end{equation}
where \(p(k_y)\) and \(q(k_x)\) are \(3\times3\) matrices:
\begin{equation*}
	\setlength{\arraycolsep}{2.25pt}
	\begin{aligned}
		p(k_y)
		&=
		\begin{bmatrix}
			0 & J_1e^{-ik_y} & 0\\
			J_1e^{ik_y} & 0 & 0\\
			0 & 0 & 0
		\end{bmatrix},
		&
		q(k_x)
		&=
		\begin{bmatrix}
			t_1 & 0 & t_2\\
			0 & t_2 & t_1\\
			t_2 & t_1 & J_2e^{ik_x}
		\end{bmatrix}.
	\end{aligned}
\end{equation*}
The perturbation that breaks \(\Gamma\) and \(M_x\) individually, while preserving their combination \(M_x\Gamma\), takes the form
\begin{equation}
	V
	=
	v\tau_3\otimes
	\begin{bmatrix}
		0 & i & 0\\
		-i & 0 & 0\\
		0 & 0 & 0
	\end{bmatrix}
	+m\tau_3\otimes 1_3,
\end{equation}
where \(\tau_i\) are the Pauli matrices. Thus, the total Hamiltonian becomes \[
\mathcal{H}(\bm{k})=\mathcal{H}_0(\bm{k})+V,
\]
and only preserves $M_x\Gamma$, as illustrated in Fig.~\ref{fig:sho-model}a.

For Fig.~\ref{fig:sho-model}c and d, the parameters are \(t_1=1.0\), \(t_2=0.8\), \(J_1=1.0\), \(J_2=2.0\), and \(v=m=0.25\).

\bigskip
\noindent \textbf{\large DATA AVAILABILITY}\\
The data generated and analyzed during this study are available from the corresponding author upon request.

\bigskip
\noindent \textbf{\large CODE AVAILABILITY}\\
All code used to generate the plotted band structures is available from the corresponding author upon request.

\bigskip
\def\bibsection{\ }
\noindent \textbf{REFERENCES}
\bibliographystyle{naturemag}

\bigskip

\noindent \textbf{\large ACKNOWLEDGEMENTS}\\
This work was supported by the Research Grants Council of Hong Kong through the General Research Fund (Grant Nos.~17302525 and 17301224).

\bigskip
\noindent \textbf{\large AUTHOR CONTRIBUTIONS}\\
R.X. and Y.X.Z. conceived the idea. Y.X.Z. supervised the project. R.X. and Y.X.Z. did the theoretical analysis. R.X. and Y.X.Z. wrote the manuscript.

\bigskip
\noindent \textbf{\large COMPETING INTERESTS}\\
The authors declare no competing interests.

\widetext
\clearpage

\includepdf[pages=1]{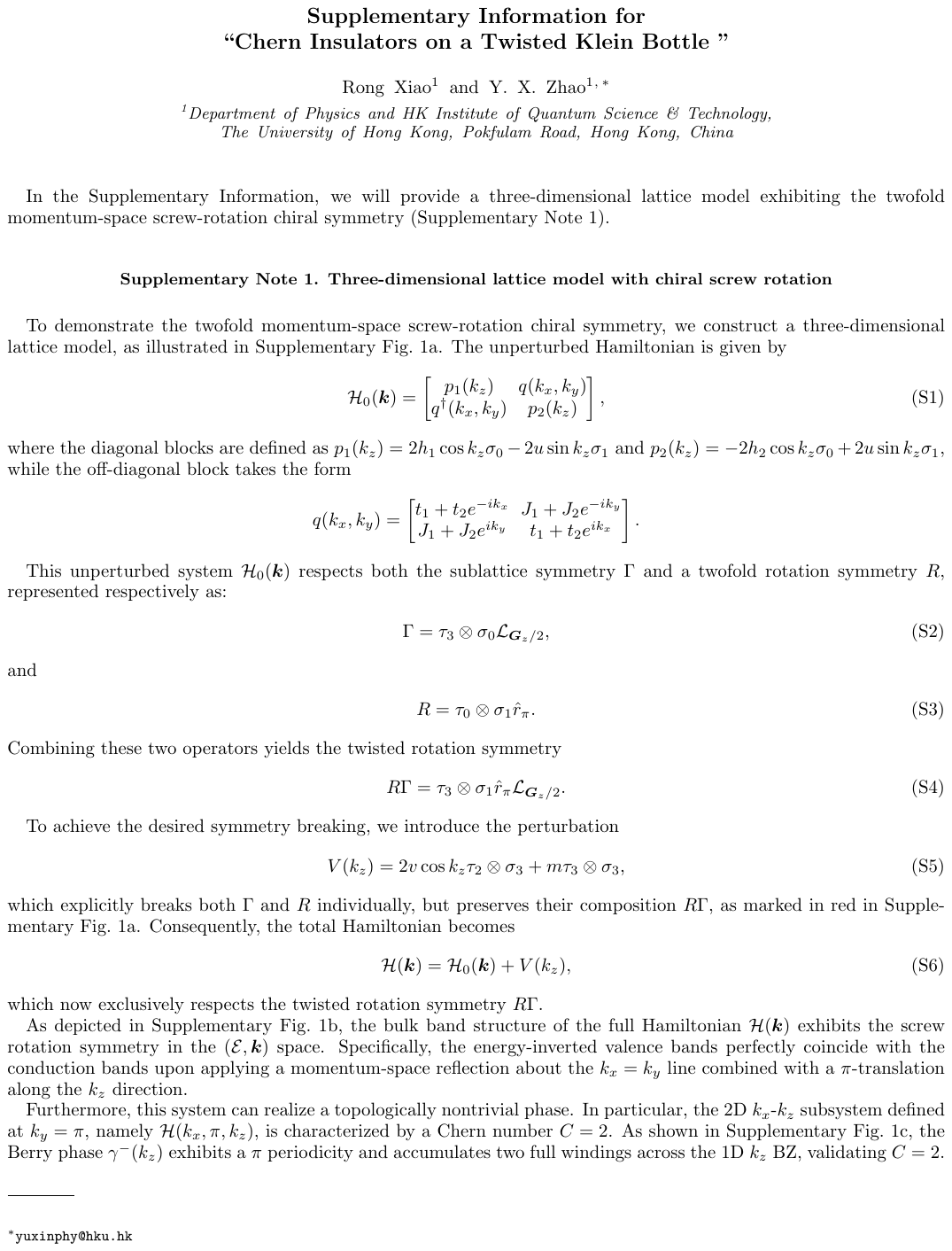}
\includepdf[pages=2]{supplement}

\end{document}